\documentclass[12pt,journal,draftcls,onecolumn,letterpaper]{IEEEtran} 

\pdfoutput=1

\usepackage{cite}

\ifCLASSINFOpdf
  \usepackage[pdftex]{graphicx}
  \graphicspath{{figures/}}
\else
\fi

\usepackage[cmex10]{amsmath}
\usepackage{array}

\usepackage{mdwmath}
\usepackage{mdwtab}

\usepackage[tight,footnotesize]{subfigure}

\usepackage[font=footnotesize]{subfig}
\usepackage{fixltx2e}

\ifCLASSOPTIONcaptionsoff
  \usepackage[nomarkers]{endfloat}
 \let\MYoriglatexcaption\caption
 \renewcommand{\caption}[2][\relax]{\MYoriglatexcaption[#2]{#2}}
\fi
\usepackage{url}

\usepackage{dsfont}
\usepackage{yfonts}

\usepackage{amsfonts}
\usepackage{times}
\usepackage{latexsym}
\usepackage{amssymb}
\usepackage{amsmath}
\usepackage{cite}

\newtheorem{theorem}{Theorem}

\newtheorem{lemma}[theorem]{Lemma}

\newtheorem{property}[theorem]{Property}

\newcommand{\figref}[1]{{Fig.}~\ref{#1}}
\newcommand{\tabref}[1]{{Table}~\ref{#1}}

\newcommand{\bI}{\mathbf{I}}
\newcommand{\cF}{\mathcal{F}}

\newcommand{\Mt}{M_t}
\newcommand{\Mr}{M_r}

\newcommand{\cN}{\mathcal{N}}

\newcommand{\Ms}{M_s}

\def\bb0{{\mathbb{0}}}

\def\bb{{\mathbf{b}}}

\def\bff{{\mathbf{f}}}

\def\bn{{\mathbf{n}}}

\def\bs{{\mathbf{s}}}

\def\bu{{\mathbf{u}}}
\def\bv{{\mathbf{v}}}
\def\bw{{\mathbf{w}}}
\def\bx{{\mathbf{x}}}
\def\by{{\mathbf{y}}}

\def\b0{{\mathbf{0}}}

\def\bD{{\mathbf{D}}}
\def\bE{{\mathbf{E}}}
\def\bF{{\mathbf{F}}}
\def\bG{{\mathbf{G}}}
\def\bH{{\mathbf{H}}}
\def\bI{{\mathbf{I}}}

\def\bS{{\mathbf{S}}}

\def\bU{{\mathbf{U}}}
\def\bV{{\mathbf{V}}}
\def\bW{{\mathbf{W}}}

\def\bbC{{\mathbb{C}}}

\def\bbF{{\mathbb{F}}}

\def\bbQ{{\mathbb{Q}}}
\def\bbR{{\mathbb{R}}}

\def\bbZ{{\mathbb{Z}}}

\def\cC{\mathcal{C}}

\def\cE{\mathcal{E}}
\def\cF{\mathcal{F}}

\def\cN{\mathcal{N}}

\def\cS{\mathcal{S}}

\def\sf0{{\mathsf{0}}}

\begin{document}

%
\title{Kerdock Codes for Limited Feedback Precoded MIMO Systems}
%
%
%

\author{Takao~Inoue,~\IEEEmembership{Member,~IEEE,}
        Robert~W.~Heath,~Jr.,~\IEEEmembership{Senior~Member,~IEEE}%
\thanks{This material is based in part upon work supported by the National Science
        Foundation under grant CCF-514194 and CNS-626797.

	Both authors are with the The University of Texas at Austin,
        Department of Electrical and Computer Engineering, Wireless
        Networking and Communication Group, 1 University Station
        C0803, Austin, TX, 78712-0240 USA. Email: \{
        inoue, rheath \} @ece.utexas.edu.}
}

%
%

\markboth{Submitted to IEEE Trans. Signal Processing, Nov. 4, 2007}%
{Inoue \& Heath: Kerdock Codes for Limited Feedback Precoded MIMO Systems}

%



\maketitle

\begin{abstract}
\boldmath
A codebook based limited feedback strategy is a practical way to
obtain partial channel state information at the transmitter in a precoded multiple-input
multiple-output (MIMO) wireless system. Conventional codebook designs
use Grassmannian packing, equiangular frames, vector
quantization, or Fourier based constructions. While the
capacity and error rate performance of conventional codebook
constructions have been extensively investigated, constructing these codebooks is notoriously difficult relying on techniques
such as nonlinear search or iterative algorithms. Further, the resulting codebooks may not have a systematic structure to facilitate storage of the codebook and low search complexity. In this paper, we propose
a new systematic codebook design based on Kerdock codes and mutually unbiased bases. The proposed Kerdock codebook consists of multiple mutually unbiased unitary bases matrices with quaternary entries and the identity matrix. We propose to derive the beamforming and precoding codebooks from this base codebook, eliminating the requirement to store multiple codebooks. The propose structure requires little memory to store and, as we show, the quaternary structure facilitates codeword search. We derive the chordal distance for two antenna and four antenna codebooks, showing that the proposed codebooks compare favorably with prior designs. Monte Carlo simulations are used to compare achievable rates and error rates for different codebooks sizes. 
\end{abstract}

\begin{IEEEkeywords}
Array signal processing, MIMO systems, feedback communication, codes.
\end{IEEEkeywords}

%
\IEEEpeerreviewmaketitle

\section{Introduction}
%
%
%
%
Multiple-input multiple-output (MIMO) systems can provide
considerable capacity and resilience to channel fading,
especially if channel state information is available at the
transmitter (CSIT) \cite{Goldsmith2003}. 
A practical solution to provide channel state information (CSI) to the transmitter is a codebook based feedback strategy, known as limited feedback or
finite rate feedback \cite{Narula1998,Mukkavilli2003,Love2003,Roh2004,Love2005a}. 
Using limited feedback, the receiver computes the appropriate
transmit precoder from a finite set of precoders, called the {\it
  codebook}, shared by the transmitter and the receiver. The receiver
then sends the index of the codeword (often in the order of few bits)
back to the transmitter resulting in the feedback of quantized
CSI. The amount of feedback depends on the size of the codebook and
the feedback rate. The size of the codebook severely
affects the system performance \cite{Narula1998,Mondal2006a}. 
Smaller codebooks reduce the feedback bandwidth but result in coarse quantization and reduced system
performance. Larger codebooks approach the ideal
CSIT case, though at an exponentially decaying rate ({\it i.e.} diminishing returns)
\cite{Mondal2006a}. In practice, $3$ to $6$ bits of feedback are
common to balance the performance and feedback tradeoff
\cite{3GPP}. A significant practical issue with limited feedback is the storage and search complexity associated with the codebook 
\cite{Ryan2007a}. Without special structure, multiple codebooks must be stored element by element with different codebooks for
beamforming and spatial multiplexing with different numbers of streams. Codeword search at the receiver grows linearly with codebook
size. In this paper, we propose a systematically constructed finite alphabet codebook that yields favorable system performance with
additional benefit of reduced storage and search complexity. 

\subsection{Background}
One of the first approaches for codebook design suggested by Narula {\it et al.} \cite{Narula1998}
is inspired by the Lloyd algorithm from
the vector quantization (VQ) literature \cite{Gersho1991}. VQ based codebook design methods have
subsequently been extensively investigated in \cite{Roh2006,Roh2006a,Xia2006,Zheng2007}.  
Roh and Rao \cite{Roh2006} proposed a mean-squared weighted inner
product (MSwIP) criteria to obtain a closed form centroid solution and 
find optimal codebooks for multiple-input single-output (MISO)
system. They extended the results to MIMO systems in \cite{Roh2006a} to
obtain codebooks that minimize capacity loss. A variation of the VQ method
by constraining the quantization space to unit hypersphere
(i.e. sphere vector quantization) was proposed by Xia and Giannakis \cite{Xia2006}. 
The VQ based approach with the Lloyd algorithm has been shown to provide
asymptotic improvement in capacity loss with increasing codebook size~\cite{Mukkavilli2003,Lau2004,Roh2004,Santipach2004,Zhou2005}.
The Lloyd algorithm requires a large number of iterations that
increase with the number of transmit antennas. Thus, VQ is primarily suitable for offline design and analytical comparisons. Furthermore,
the VQ based codebook does not yield any structure in the
codebook. Each entry, often complex valued, with maximum available
precision must be stored. For codebook selection at the receiver, an
exhaustive search with complex matrix operation is usually
required \cite{Ryan2007a}. 

The Grassmannian packing approach for codebook design was
proposed for beamforming by Love {\it et al.} \cite{Love2003} and Mukkavilli {\it et al.} \cite{Mukkavilli2003}, then later extended to precoding by Love and Heath \cite{Love2005a}. The problem of Grassmannian packing is to find a set of subspaces, which are points in the Grassmann manifold, and are maximally spaced in terms of subspace distance. Quantization on the Grassmann manifold has been studied by Mondal
{\it et al.} \cite{Mondal2006a,Mondal2007}. A major challenge associated with Grassmannian codebooks is that it is difficult to find good packings and thus good codebooks. The real case has been studied from an algebraic point of view by Conway {\it et al.} \cite{Conway1996} and Shor and Sloane \cite{Shor1998}, 
while bounds on maximum minimum distance were investigated by Bachoc \cite{Bachoc2006} and Barg and Nogin
\cite{Barg2006}.  An extensive computer search is usually required
(i.e. offline design) and the resulting codebook generally
does not have any structure to ease the storage and search computation 
requirement \cite{Tropp2006}.

Mondal {\it et al.} proposed an equiangular
frame (EF) based codebook using mutual information criteria and showed
that (at least in the real case) the EF codebook maximizes the packing radius if and only if it is
an EF, and that EF codebooks achieve lower average distortion compared
to VQ or Grassmannian codebooks \cite{Mondal2005a}. The EF codebook is based on
Grassmannian frames proposed by Strohmer and Heath \cite{Strohmer2003}, and are also connected with Grassmannian packings, spherical designs, and regular graphs. As with Grassmannian packing, constructing equiangular frames is also challenging. 
Tropp {\it et al.} proposed an alternating projection method to
construct equiangular frames \cite{Tropp2005}. Alternating
projection algorithm iteratively projects onto two constraint sets,
a unit norm constraint set and unit tight frame constraint set, to
arrive at the approximate equiangular frame with prescribed error. 
Unfortunately, apart from the equiangular distance property,
the resulting codebook does not have any constructive
structure and the alternating projection algorithm does not work well
for small finite alphabets \cite{Tropp2005}. 

One popular codebook is the Fourier codebook proposed in \cite{Love2003,Love2005a}, inspired by 
the unitary space-time constellation design in Hochwald {\it et al.} \cite{Hochwald2000}. Exhaustive search is used to find a set of Fourier matrices, and modifications of Fourier matrices by a diagonal generator matrix of complex exponentials that maximize the minimum correlation between codewords.  This construction is being considered for 3GPP long-term evolution
(LTE) and 3GPP2 ultra mobile broadband (UMB) due to its systematic
codebook generation \cite{3GPP}. A connection between the Fourier based design, the Welch
bound \cite{Welch1974}, and difference sets were reported in
\cite{Xia2005}. For practical applications, the Fourier based design
is attractive because only two matrices, the generator matrix and the
discrete Fourier transform (DFT) matrix, need to be stored. Codeword search is somewhat simplified at the receiver since it can exploit the structure of the DFT matrix, but matrix computations
with complex arithmetic are still required because the codebook contains
complex values.  

To minimize storage requirements and search complexity, 
a quadrature amplitude modulation (QAM) based codebook was proposed by
Ryan {\it et al.} \cite{Ryan2007a}. The codebook design was motivated
by limiting the codeword entries to a finite alphabet (i.e. QAM constellation
points) and reducing the search complexity using maximum likelihood
non-coherent lattice decoding \cite{Ryan2006}. This method is
essentially an element by element quantization of the precoder matrix
to the nearest QAM points. The main
drawback is that the resulting feedback rate is larger than other
codebook constructions, which makes it less attractive in practice.

\subsection{Contributions}

In this paper, we propose a single user codebook design for limited feedback MIMO systems based on Kerdock codes.  Kerdock codes were originally proposed for error correction \cite{Kerdock1972}. It was shown
by Hammons {\it et al.} \cite{Hammons1994} that Kerdock codes are
$\bbZ_4$ (integer modulo $4$ or quaternary alphabet) linear containing
more codewords than any known linear code with the same minimal
distance. We exploit the connection between Kerdock codes and mutually
unbiased bases (see e.g. \cite{Heath2006} for details). 
Our proposed Kerdock codebook consists of multiple mutually unbiased
unitary bases matrices with quaternary entries and the identity
matrix. This  codebook is used to derive codebooks for beamforming and
precoding. The beamforming codebook is derived from all the columns of
the codebook while the precoding codebooks are derived by subsets of
columns from each bases. Note that the inclusion of the identity
matrix, which is not ad hoc but actually forms one of the mutually
unbiased bases, means that antenna subset selection \cite{Heath2001} is
included as a special case. We consider two practical examples of
codebooks for two and four transmit antennas using two different
constructions: a Sylvester-Hadamard construction \cite{Heath2006} and
a power construction \cite{Gow2007}. The Sylvester-Hadamard
construction gives a good solution for the two antenna case while the
structure in the power construction gives a better solution for the
four antenna case. The power construction also facilitates closed form
derivation of certain subspace distance properties. While we are
mainly concerned with narrowband single user MIMO system in this
paper, multiuser MIMO \cite{Spencer2004a} and wideband systems \cite{Choi2005} appears to be promising.

The application of Kerdock codes is motivated by the subspace distance property with the additional practical benefit of
storage and search complexity reduction. We demonstrate the benefits of  the proposed Kerdock codebook using the following metrics:
\begin{enumerate}
  \item system performance,
    \item proximity to a Grassmannian codebook,
  \item construction \& storage,
  \item search efficiency.
\end{enumerate}
To evaluate system performance, we compare the vector symbol error rate (VSER) and the achievable rate with different codebooks and with perfect CSIT. Our Monte Carlo simulation results show that Kerdock codebooks have negligible performance loss, and in some cases performance gain, with previously proposed codebooks. The proposed Kerdock codebook satisfies the sufficient conditions in \cite{Mondal2006c}, and thus is full-diversity. To measure proximity to a Grassmannian codebook, we compute the maximum minimum subspace distance (in this case we use the Fubini-Study distance) and compare with the best known Grassmannian packings. Because of the special structure of our codebook, we are able to derive exactly the Fubini-study distance for our two and four antenna codebooks for all dimensions of precoders in closed form. This special structure shows that the proposed Kerdock codebooks are quite close to Grassmannian codebooks. To evaluate construction and storage, we estimate the required storage as a function of the number of bits of precision including storing multiple codebooks for different modes of transmission  ({\it i.e.} beamforming and spatial multiplexing). The storage required for Kerdock codes is much smaller than for general Grassmannian, VQ, or EF codebooks and is somewhat smaller than the Fourier construction due to the quaternary structure. Finally, we estimate the search efficiency by estimating the number of operations required to find the optimum codeword in the codebook at the receiver. Here the main simplification comes from the fact that the entries of the codebook are either scaled version of $\{1,-1,j,-j\}$ or are zero. Consequently the multiplication operations become simple sign flipping or flipping real and imaginary components. 

\subsection{Organization}
This paper is organized as follows. In Section
\ref{sec:system}, an overview of the system model is given. In
Section \ref{sec:mub}, the construction of the MUB codebook is
outlined and the strategies for precoded MIMO systems are given. In
Section \ref{sec:storage}, we analyze the storage and search
complexity compared with Grassmannian and Fourier based design.
In Section \ref{sec:analysis}, we analyze the distance, diversity, and capacity
performance. In Section \ref{sec:simulation}, we provide numerical
simulation results to support our analysis. Finally, we conclude this
paper with some remarks in Section \ref{sec:conclusion}.

For notation, we use lower case bold letters, {\it e.g.} ${\bf v}$,
to denote vectors and upper case bold letters, {\it e.g.} ${\bf H}$, to
denote matrices. The $n \times n$ identity matrix is denoted by
$\bI_n$. The space of real and complex are denoted by $\bbR$ and
$\bbC$, respectively with an appropriate superscript to denote the
dimension of the respective spaces. We shall use $^T$ and $^*$ to
denote the transposition and Hermitian transpose, respectively.

\section{System Overview}\label{sec:system}

\subsection{Discrete-time System Model}
\begin{figure}[ht]
\centering
\includegraphics[width=5in]{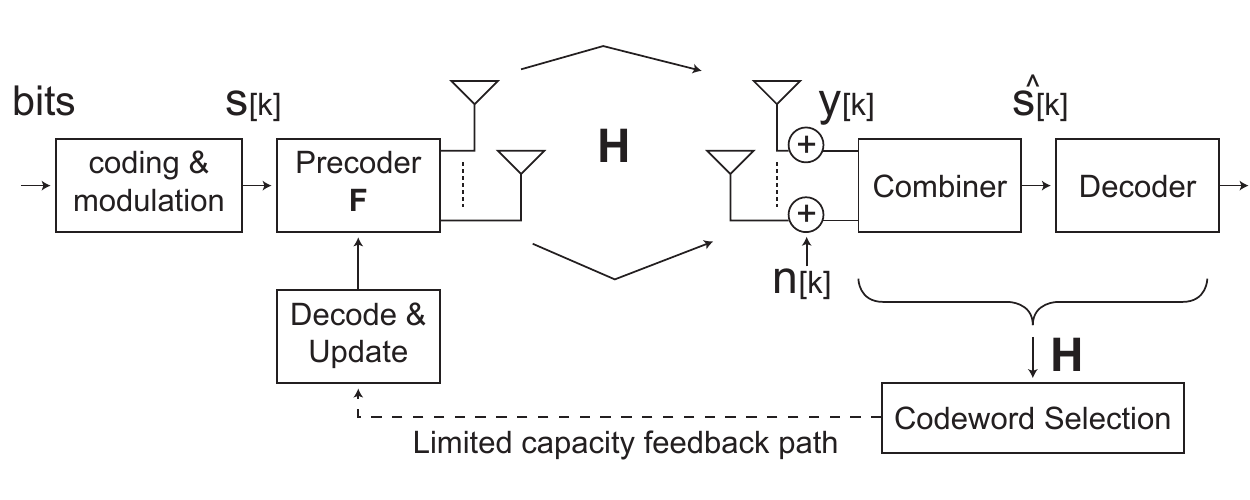}
\caption{Block diagram of general limited feedback MIMO System}
\label{fig:General_LFMIMO}
\end{figure}

A limited feedback precoded MIMO wireless system with $\Mt$
transmit antennas and $\Mr$ receive antennas is shown in
\figref{fig:General_LFMIMO}.
The transmit bit stream is sent to the encoder and modulator, which
outputs a complex transmit vector, $\bs[k] = [ s_{1}[k], s_{2}[k],
\dots, s_{\Ms}[k] ]^T$, where $k$ denotes the time index and $\Ms$
denotes the number of streams to be sent. Note that beamforming is
the special case where $\Ms = 1$, and $1 < \Ms \leq \Mt$ for
$\Ms$-stream spatial
multiplexing. We assume that $E_{\bs} [\bs \bs^*] = \frac{\cE_s}{\Ms}\bI_{\Ms}$
to constrain the average transmit power, $E_{\bs}$ is used to
denote the expectation with respect to the transmit vector, and $\cE_s$
is used to denote the total transmit power.

The transmit vector $\bs[k]$ is then multiplied by the unitary precoder
$\bF$ ($\bff$ for beamforming) of size $\Mt \times \Ms$ with 
$\bF^* \bF = (1/\Ms)\bI_{\Ms}$, producing a length $\Mt$ transmit vector 
$\bx[k] = \sqrt{\cE_s / \Ms}\bF \bs[k]$. The precoder $\bF$ is
selected based on limited feedback information from the receiver. 

Assuming perfect synchronization, sampling, and a linear memoryless
channel, the equivalent baseband input-output relationship can be
written as
\begin{equation}
  \by[k] = \sqrt{\frac{\cE_s}{\Ms}} \bH \bF \bs[k] + \bn[k]
  \label{eq:GeneralInputOutput}
\end{equation}
where $\bH$ is the channel matrix and $\bn[k]$ is the noise vector. We
assume that $\bH$ is an $\Mr \times \Mt$ matrix where each entry is
unit variance complex Gaussian independent and identically distributed (i.i.d.) according to $\cC\cN(0,1)$.
The entries of $\bn[k]$ are also complex Gaussian i.i.d. distribution according to
$\cC\cN(0,N_0)$. The receive vector $\by[k]$ is then decoded by assuming
a perfect knowledge of $\bH \bF$ at the receiver to produce the output
vector $\hat{\bs}$.

We assume that the receiver has a perfect estimate of the channel
matrix $\bH$ and uses a linear receiver which applies an $\Ms \times \Mr$ matrix
$\bG$ to the receive symbol $\by[k]$. For the spatial multiplexing case,
the zero-forcing (ZF) receiver is used which is given by 
$\bG = (\bH \bF)^{\dagger}$ where $(\cdot)^{\dagger}$ denotes the
Moore-Penrose pseudo inverse. 
For the beamforming case, a maximum ratio combining (MRC) receiver is
assumed with $\bG = (\bH\bF)^*$. 

\subsection{Codeword Search and Selection}
In this paper, the receiver chooses the precoding matrix
(codeword) $\bF$ from a finite set of $N$ possible codewords $\cF = 
\{ \bF_1, \bF_2, \dots, \bF_N \}$ called the {\it codebook} shared by the
transmitter and the receiver. The index of the selected codeword at
the receiver, based on the knowledge of the channel, is fed back to
the transmitter through a zero-delay limited capacity feedback
channel. The codeword index $n=1,2,\dots,N$ is represented by $b$-bit
binary ($N=2^b$) resulting in $b$ bits of feedback. We say that
the codebook is {\it $b$-bit codebook} when it has $N =2^b$
entries.

For beamforming, the optimal beamformer 
maximizes the effective SNR \cite{Love2003}, 
\begin{equation}
  \bff = \arg \max_{\bw \in \cF} \| \bH \bw \|_2^2.
  \label{eq:BF_SC}
\end{equation}
The chordal distance between codeword vectors, $\bff_1$ and $\bff_2$,
is given by 
\begin{equation}
  d_{\text{ch}}(\bff_1,\bff_2) = \sin(\theta_{1,2}) 
  = \sqrt{1 - |\bff_1^* \bff|^2}.
  \label{eq:chordal_dist}
\end{equation}
The chordal distance is used to analyze the distance property of the
beamforming codebook. The beamformer selection can also be approximated 
by finding the beamforming vector from the codebook
$\cF$ with the minimum chordal distance, to the right singular vector,
$\bv$, corresponding to the largest singular value of the channel
$\bH$ \cite{Mondal2006c}. 

For spatial multiplexing, the optimal unitary precoder is given by 
$\bF_{\text{opt}} = [\bV]_{1:\Ms}$ where $\bV$ is the right singular
matrix of the channel $\bH$ and $[\cdot]_{1:\Ms}$ denotes the
first to $\Ms$-th columns of the given matrix \cite{Love2005a,Mondal2006c}.
We employ the minimum singular value selection criteria (MSV-SC) \cite{Love2005a}
\begin{eqnarray}
  \bF & = & \arg \max_{\bW \in \cF} \lambda_{\text{min}} \{ \bH \bW \}
  \label{eq:SM_SC}
\end{eqnarray}
where $\lambda_{\text{min}}$ denotes the minimum singular value of the
product $\bH \bW$. 
This selection criteria approximately maximizes the minimum substream SNR. 

We use the two distance criteria proposed in \cite{Love2005a} to
evaluate the codebooks . It was shown in 
\cite{Love2005a} that the codebook should be designed by maximizing
either the minimum projection 2-norm distance,
$\min_{\bF_1 \neq \bF_2} d_{p2}(\bF_1,\bF_2)$, or the minimum Fubini-study
distance, $\min_{\bF_1 \neq \bF_2} d_{FS}(\bF_1,\bF_2)$. 
The projection 2-norm distance is defined as
\begin{eqnarray}
  d_{p2}(\bF_1,\bF_2) & = & 
  \|\bF_1 \bF_1^* - \bF_2 \bF_2^* \| \nonumber \\
  & =& \sqrt{1 - \lambda_{\text{min}}\{\bF_1^*
  \bF_2\}},
  \label{eq:p2_dist}
\end{eqnarray}
and the Fubini-study distance is defined as
\begin{eqnarray}
  d_{FS}(\bF_1,\bF_2) & = &
  \text{arc} \cos | \det (\bF_1^* \bF_2)|.
  \label{eq:fs_dist}
\end{eqnarray}

In this paper, we will be concerned with selection criteria
\eqref{eq:BF_SC} for beamforming and \eqref{eq:SM_SC} for spatial
multiplexing, respectively. To analyze the distance properties of the
codebook, \eqref{eq:chordal_dist} will be analyzed for beamforming and 
\eqref{eq:p2_dist} and \eqref{eq:fs_dist} are analyzed for spatial multiplexing.

\section{Kerdock Codebook}\label{sec:mub}
In this section, we provide the background information on Kerdock codes and mutually
unbiased bases (MUB), their construction, and their utility as a limited
feedback codebook. 

\subsection{Preliminaries}\label{sec:mub:prelim}
Mutually unbiased bases (MUB) arises in
connection with quantum information theory where the observable state
of a quantum system is represented by the set of orthonormal bases
(ONB) with certain correlation property (to be defined shortly) \cite{Klappenecker2005}.
Some of the known MUB constructions are due to Alltop
\cite{Alltop1980}, Wootters and Fields \cite{Wootters1989},
Klappenecker and Roetteler \cite{Klappenecker2004}, Bandyopadhyay
{\it et al.} \cite{Bandyopadhyay2002}, and recently by Gow
\cite{Gow2007}. 
Klappenecker and Roetteler \cite{Klappenecker2005} showed that many of
the MUB constructions are equivalent and that these constructions have
a close connection with complex projective space and uniform tight
frames, both of which has been used for the construction and analysis
of quantized codebooks for limited feedback MIMO systems. 
Based on these connections, we study the utility of MUB as a limited feedback
codebook. 

An MUB is a collection of two or more ONB with
the property that the columns of different ONBs has the same
correlation (or inner product). That is, if $\bS = \begin{bmatrix} \bs_1  \dots
  \bs_{\Mt} \end{bmatrix}$ and $\bU = \begin{bmatrix} \bu_1 & \dots
  & \bu_{\Mt} \end{bmatrix}$ are two $\Mt \times \Mt$ ONBs ({\it i.e.} $\bS^* \bS =
\bI_{\Mt}$), the inner product of vectors drawn from each ONB
satisfies the {\it mutually unbiased} property
\begin{equation}
  | \langle \bs_n, \bu_m \rangle | = \frac{1}{\sqrt{\Mt}}
\end{equation}
for $n,m=1,\dots,\Mt$. MUB is the set $\cS = \{\bS_0, \bS_1,
\dots, \bS_{N_s}\}$ with each ONB $\bS_n$, $n=0,\dots,N_s$, satisfying the
mutually unbiased property. Two natural questions to ask are 1) how
many such bases exists for a given dimension $\Mt$, and 2) how to
construct the MUB. First, the maximum number of ONBs, $N_s = |\cS|$, has been shown
to be $N_s \leq \Mt + 1$ with equality if $\Mt$ is a power of a
prime \cite{Bandyopadhyay2002,Klappenecker2004}. It is presently unknown whether
equality occurs when $\Mt$ is not a power of prime and this question
remains to be an active area of research \cite{Gow2007}. Second,
several approaches for the construction of size $\Mt+1$ MUB for prime
powers has been proposed (see \cite{Planat2006} for an excellent
survey and references therein). In this paper, we shall be concerned
with power of two construction, mainly for $\Mt=2$ and $4$, for its
attractive construction and practical applicability. One drawback is
that we do not have a construction for $\Mt=3$ with the finite
alphabet properties so we shall differ this case for future work. 

We now turn our attention to why MUB is a useful limited feedback
codebook design. To see this, consider the beamforming codebook
$\{\bff_i\}_{i=1}^N \in \cF$. The defining characteristics of MUB is 
its mutually unbiased property. That is, for $k,l =1,
\dots, N$, $|f_k^* f_l | = 0$ if $\bff_k$ and $\bff_l$ are chosen from the same
bases and $|f_k^* f_l | = 1/\sqrt{\Mt}$ if $\bff_k$ and $\bff_l$ are
chosen from a different bases. We can compute the average inner
product of the codebook as
\begin{equation}
\frac{1}{N(N-1)} \sum_{l=1}^N \sum_{l' \neq l}^N | \bff_l^*
\bff_{l'}|^2 = 
\frac{N-\Mt}{(N-1)\Mt}.
\label{eq:ave_dist}
\end{equation}
The Grassmannian packing problem is to maximize the minimum pairwise
distance of codewords using the distance function
\eqref{eq:chordal_dist}. Substituting \eqref{eq:ave_dist} as the average
inner product in \eqref{eq:chordal_dist}, we obtain an approximate
distance bound of the codebook
\begin{eqnarray}
  d_{\text{ch}}(\cF) & \approx & \sqrt{1 - \frac{N-\Mt}{(N-1)\Mt}} \nonumber
  \\
  & = & \sqrt{\frac{N(\Mt-1)}{\Mt(N-1)}}
\end{eqnarray}
which is the same as the well known Rankin bound, an upper bound on
the minimum distance for line packings \cite{Love2003}. Therefore, the
MUB codebook is near optimal in the sense that the Rankin
bound is met in an average sense. 

\subsection{Sylvester-Hadamard Codebook Construction}\label{sec:mub:kerdock}
Kerdock codes originally proposed for error correcting
codes \cite{Hammons1994} are known to be MUB \cite{Planat2006}. MUB
structure of Kerdock codes were used to design a scalable signature
sequence for code division multiple access (CDMA) system by Heath 
{\it et al.} \cite{Heath2006}. Kerdock codes are particularly
attractive for implementation since the codebook contains finite
alphabet, $\{\pm 1, \pm j\}$, and still satisfy the mutually unbiased
property. We shall use the simplified Kerdock construction in
\cite{Heath2006} to construct $\Mt = 2$ finite alphabet codebook. For
$\Mt=4$, we first construct the Kerdock code and modify it to obtain
the power construction described in Section
\ref{sec:mub:construction}. 

The Kerdock code construction proposed in \cite{Heath2006} consists of
$\Mt=2^B$ orthonormal matrices, where $B$ is a positive integer. 
Orthonormal matrices are denoted $\bS_n$, $n=0,\dots,\Mt-1$, where each
$\bS_n$ is a rotated Sylvester-Hadamard matrix. The key to the
construction is the algebraic derivation of
the rotating (or generator) matrices, $\bD_n$, which does not rely on
any search.

Let $\hat{\bH}_{\Mt}$ denote the size $\Mt \times \Mt$
Sylvester-Hadamard matrix where
\begin{equation}
  \hat{\bH}_2 = \begin{bmatrix} 1 & 1 \\ 1 & -1 \end{bmatrix}
  \label{eq:hadamard}
\end{equation}
and 
\begin{equation}
  \hat{\bH}_{\Mt} = \underbrace{\hat{\bH}_2 \otimes \hat{\bH}_2
  \cdots}_{B \text{times}}.
\end{equation}
The general strategy for the Kerdock codebook construction is as follows:
\begin{enumerate}
  \item Construct the diagonal matrices, $\bD_n$, for
  $n=0,1,\dots,\Mt-1$. These are the generator matrices.
  \item Each bases is constructed by $\bS_n = (1/\sqrt{\Mt})\bD_n \hat{\bH}_{\Mt}$. 
  \item Let $\hat{\bF} = [ \bS_0 \bS_1 \cdots \bS_{\Mt-1}]$.
\end{enumerate}
For brevity, we omit the details of the diagonal generator matrix
construction which can be found in \cite{Heath2006}. 

\subsection{Power Codebook Construction}\label{sec:mub:construction}
Let $p$ be a prime number and let $q=p^a$, where $a$ is a
positive integer. Let $\bbF$ denote the finite field of order
$q^2$. We let $G_q$ denote the finite group $\bbF \times \bbF$ of
order $q^4$ with multiplication defined by 
\begin{equation}
  (a,b)(c,d) = (a+c,a^qc + b + d).
\end{equation}

\begin{lemma}
Let $\alpha$ be an element of order $q+1$ in $\bbF$. The the mapping
$\sigma: G_q \to G_q$ given by $\sigma(a,b)=(\alpha a,b)$ is an
automorphism of order $q+1$ of $G_q$. 
\end{lemma}

Let $\chi$ be an irreducible character of $G_q$ of degree $q$ and $X$
be an irreducible representation of $G_q$ with character $\chi$. 
The key result due to Gow \cite{Gow2007} is the following.
\begin{theorem}
  Let $p=2$ and let $X$ be of degree $q$ consisting of unitary
  matrices. If $D$ is a $q \times q$ matrix that satisfies $\bD^{q+1}
  = \bI$ and $\bD^{-1}X(x)\bD = X(\sigma(x))$ for all $x$ in
  $G_q$, then the powers $\bD, \bD^2, \dots, \bD^{q+1} = \bI$ generates
  $q+1$ pairwise mutually unbiased bases. Furthermore, all entries of
  $\bD$ are in the field $\bbQ(\sqrt{-1})$. 
  \label{theorem:MUBpower}
\end{theorem}
Theorem \ref{theorem:MUBpower} states that if a matrix $\bD$ which
satisfies the given conditions is found, then the powers of $\bD$
generates the size $q+1$ MUB. 

For the limited feedback codebook design, Theorem
\ref{theorem:MUBpower} represents a powerful result when
the number of transmit antennas are power of 2. Only the generating
base $\bD$ needs to be stored and the rest of the codebook can be
generated from the products. Therefore
\begin{eqnarray}
  \cS = \{ \bS_0 = \bD, \bS_1 = \bD^2, \dots, \bS_{N_s-1} = \bD^{N_s}
  \}.
  \label{eq:MUBpower}
\end{eqnarray}
Note also the inclusion of identity element which corresponds to the
case of antenna subset selection \cite{Heath2001}. All of the previously proposed codebooks do
not have identity as part of the codebook. In standards such as 3GPP LTE,
the identity element is included in the codebook \cite{3GPP} for
antenna selection. The MUB construction extends naturally for use in
3GPP LTE. 

\subsection{Codebook Arrangement}\label{sec:mub:arrange}
Once the MUB is generated, we apply the following procedure to arrange $\bS_n$ into
a codebook. For the beamforming system, we construct the composite matrix,
$\hat{\bF} = \begin{bmatrix} \bS_0 & \bS_1 & \cdots & \bS_{N_s -1}
\end{bmatrix}$ and define the codebook as the columns of
$\hat{\bF}$. That is
\begin{equation}
  \cF = \{ \bF_1 = [\hat{\bF}]_1, \bF_2 = [\hat{\bF}]_2, \dots,
  \bF_N = [\hat{\bF}]_N \}
  \label{eq:KBFCodebook}
\end{equation}
where $[ \cdot ]_n$ is used to denote the selection of $n$-th column
of a given matrix and $N \leq \Mt N_s$. 

For the unitary precoding spatial multiplexing system, a subset of
columns are selected from each $\bS_n$ to form the codebook. Notice that
for a subset of columns drawn from a single $\bS_n$, each columns are
orthogonal to each other by construction. Hence, the design problem is to
take a column subset from each $\bS_n$ so that the pairwise minimum distance,
\eqref{eq:p2_dist} or \eqref{eq:fs_dist}, is maximized and to see whether
such codebooks can yield performance similar to previously proposed
codebooks.

For $\Ms$-stream spatial multiplexing codebook, take all $\Ms$-column
subsets from each $\bS_n$. There are $\begin{pmatrix}\Mt \\ \Ms
\end{pmatrix}$ column subset combinations in each $\bS_n$. The maximum
number of codewords that the MUB can take is 
$N_s \times \begin{pmatrix}\Mt \\ \Ms \end{pmatrix}$. We shall
illustrate the system performance for a few possible subsets of so
constructed codebook in Section \ref{sec:simulation}. 
One way to select the subset is by identifying a unique column
combinations from each $\bS_n$ so that every pairwise minimum distance 
(either \eqref{eq:p2_dist} or \eqref{eq:fs_dist}) is maximized.
Unfortunately, exhaustive search appears to be the only way to find
the best combination. 

In Section \ref{sec:analysis}, we show that constant distance between
codewords can be achieved by this construction, and
in Section \ref{sec:simulation}, we show through numerical simulation
that this codebook performs comparable to same sized Grassmannian and
Fourier based codebooks. We do not claim optimality of the
so constructed spatial multiplexing codebook other than that it
achieves full diversity.  

\subsection{Two Transmit Antenna Construction}
In this section, we provide an example of Kerdock codebook
construction for two antenna MIMO system. 
This is the trivial case for the Kerdock codebook
construction and it is easy to verify that 
\begin{equation}
  \bD_0 = \begin{bmatrix} 1 & 0 \\ 0 & 1 \end{bmatrix},
  \bD_1 = \begin{bmatrix} 1 & 0 \\ 0 & j \end{bmatrix}.
\end{equation}
The resulting $\bS_n$ are
\begin{equation}
  \bS_0 = \frac{1}{\sqrt{2}}\begin{bmatrix} 1 & 1 \\ 1 & -1 \end{bmatrix},
  \bS_1 = \frac{1}{\sqrt{2}}\begin{bmatrix} 1 & 1 \\ j & -j
  \end{bmatrix},
  \bS_2 = \begin{bmatrix} 1 & 0 \\ 0 & 1 \end{bmatrix},
\end{equation}
where $\bS_0$ is merely the scaled Sylvester-Hadamard matrix,
$\hat{\bH}_2$, with quaternary alphabet.

In this case we prefer to to use the Sylvester-Hadamard construction since the power construction for $\Mt=2$ is
\begin{eqnarray}
  \bD & = & \frac{1}{2}\begin{bmatrix} -1-j & -1+j \\ 1+j & -1+j \end{bmatrix}
\end{eqnarray}
is not exactly quaternary. Taking the powers of $\bD$, we readily obtain the following MUB
\begin{eqnarray}
  \left\{ 
  \frac{1}{2}\begin{bmatrix} -1-j & -1+j \\ 1+j & -1+j \end{bmatrix},
  \frac{1}{2}\begin{bmatrix} -1+j & 1-j \\ -1-j & -1-j \end{bmatrix},
  \begin{bmatrix} 1 & 0 \\ 0 & 1 \end{bmatrix}
  \right \}.
\end{eqnarray}
This codebook does have the finite alphabet property and also includes the identity
corresponding to antenna selection.  

The beamforming codebook is constructed by forming the composite matrix $\hat{\bF} =
\begin{bmatrix} \bS_0 & \bS_1 & \bS_2 \end{bmatrix}$ and considering each
column as beamforming vectors,
\begin{equation}
  \cF = \{ \bF_1 = [\hat{\bF}]_1, \bF_2 = [\hat{\bF}]_2, \dots,
  \bF_N = [\hat{\bF}]_6 \}.
\end{equation}

\subsection{Four Transmit Antenna Construction}
In this section, we give an example MUB codebook for $\Mt=4$. 
Our simulation results in Section \ref{sec:simulation} utilize the
codebook constructed in this example. 

For $\Mt = 4$, we start with the simplified Kerdock code construction
in \cite{Heath2006} and making a slight modification to one of the
bases, we find that the following generator matrix satisfies Theorem
\ref{theorem:MUBpower} 
\begin{eqnarray}
  \bD & = & \frac{1}{2} \begin{bmatrix}
  -j & -j & -j & -j \\
  1 & -1 & 1 & -1 \\
  -j & -j & j & j \\
  -1 & j & j & -j
  \end{bmatrix}.
\end{eqnarray}
By inspection, we can recognize that $\bD$ can be decomposed
with a Sylvester-Hadamard matrix as follows
\begin{eqnarray}
  \bD = \frac{1}{2} \begin{bmatrix}
  -j & 0 & 0 & 0 \\
  0 & 1 & 0 & 0 \\
  0 & 0 & -j & 0 \\
  0 & 0 & 0 & -1
  \end{bmatrix} 
  \left( 
  \hat{\bH}_2 \otimes \hat{\bH}_2
  \right).
  \label{eq:MUBdecomposition}
\end{eqnarray}

Finally, computing $\bS_n = \bD^{n+1}$ yields
\begin{eqnarray}
  \bS_0 & = & \frac{1}{2} \begin{bmatrix}
    -j & -j & -j & -j \\
    1 & -1 & 1 & -1 \\
    -j & -j & j & j \\
    -1 & 1 & 1 & -1
  \end{bmatrix} \nonumber \\
  \bS_1 & = & \frac{1}{2}\begin{bmatrix}
    -1 & -1 & -j & j \\
    -j & -j & -1 & 1 \\
    -j & j & -1 & -1 \\
    1 & -1 & j & j 
  \end{bmatrix} \nonumber \\
  \bS_2 & = & \frac{1}{2}\begin{bmatrix}
    -1 & j & j & 1 \\
    -1 & j & -j & -1 \\
    j & -1 & -1 & -j \\
    -j & 1 & -1 & -j
  \end{bmatrix} \nonumber \\
  \bS_3 & = & \frac{1}{2}\begin{bmatrix}
    j & 1 & j & -1 \\
    j & -1 & j & 1 \\
    j & 1 & -j & 1 \\
    j & -1 & -j & -1
  \end{bmatrix} \nonumber \\
    \bS_4 & = & \bI_4 .
\end{eqnarray}
Thus we obtain an MUB with quaternary alphabet based on Kerdock
codes. 

For the beamforming system, we construct
the composite matrix from the above construction,
$\hat{\bF} = \begin{bmatrix} \bS_0 & \bS_1 & \bS_2 & \bS_3 & \bS_4
\end{bmatrix}$ and define the codebook as the columns of
$\hat{\bF}$. That is 
\begin{equation}
  \cF = \{ \bF_1 = [\hat{\bF}]_1, \bF_2 = [\hat{\bF}]_2, \dots,
  \bF_N = [\hat{\bF}]_{20} \},
  \label{eq:KBF4}
\end{equation}
for a $N=20$, $5$-bit codebook. The identity element, $\bS_4$, can be
deleted if antenna selection is not needed. 

For the unitary precoding spatial multiplexing system, we shall take
all $\Ms$ column subset from each $\bS_n$, $n=0,1,\dots,4$. For $\Ms=2$,
we obtain $5 \times \begin{pmatrix} 4 \\ 2 \end{pmatrix} = 30$
codewords, or $5$-bit codebook, and for $\Ms=3$, $5 \times
\begin{pmatrix} 4 \\ 3 \end{pmatrix} = 20$ codewords, or $5$-bit
codebook. The distance properties of the so obtained codebooks are
analyzed in Section \ref{sec:analysis}.

We have thus obtained a finite alphabet codebook which can be shared
for beamforming and spatial multiplexing. Next, we will analyze and
quantify the storage and search complexity associated with Kerdock codes
and compare it with Grassmannian and Fourier based construction.

\section{Codebook Storage and Search Complexity}\label{sec:storage}
Previous limited feedback codebook
designs were primarily concerned with achievable performance
\cite{Love2003,Love2005a}, but the resulting codebook was such that the
codebooks of various sizes  for
different modes of transmission had to be stored at both the
transmitter and the receiver. In today's hand-held mobile devices,
power consumption and device size are two major design challenges 
\cite{Nilsson2005,Markovic2007}. A large portion of
today's baseband devices are memory elements while power consumption can
be related to the amount of computation required on the device.
To minimize the impact of codebook based limited feedback
implementation, it is of great interest to minimize the
codebook storage and search computation for various modes of transmission. 
Furthermore, in recent standards such as 3GPP LTE and 3GPP2 UMB, up
to 300km/h of mobility is being considered \cite{3GPP}. A simple search algorithm
is of great interest to reduce the time of adaptation in highly mobile
environment. In this section, we quantify the storage and search
complexity of the proposed MUB codebook and compare it with
Grassmannian and Fourier based designs. 

\subsection{Storage}
To estimate the storage requirements, we will consider the number of real elements
({\it i.e.} two real components for one complex value) to store a
codebook for each mode of transmission ({\it i.e.} beamforming and
spatial multiplexing). Let $N_b$ denote the number of bits available
in the system to represent a real number.
It is easy to see that the storage
required for a single codebook with $N$ entries of $\Mt \times \Ms$
complex-entry codewords for a particular transmission mode is upper
bounded by $2 N_b N \Mt \Ms$-bits. Note that some reduction in number
of stored bits may be possible due to specific values taken on by the
codeword entries, but we will only contend with the worse case scenario. 

The Grassmannian codebook \cite{Love2003,Love2005a,LoveCodebook} does not yield any systematic construction
so the entire codebook, element by element, must be stored. Thus, the
storage requirement is $2 N_b N \Mt \Ms$-bits for each codebook. 

The Fourier based codebook \cite{Hochwald2000} requires the storage of one diagonal
generator matrix ({\it i.e.} $\Mt$ complex entries) and $\Mt \times \Ms$
entries of a DFT matrix. In general, the generator matrix is
different for each mode of transmission but the $\Mt \times \Mt$
DFT matrix can be used for all cases. The storage requirement for
Fourier based codebook is $2 N_b (\Mt + \Mt \Ms)$-bits for a given
transmission mode. Note that the storage requirement is independent of
the codebook size, $N$, because the generator matrix is designed for a
particular codebook size $N$. 

The MUB codebook construction in Section \ref{sec:mub}
requires storage of only the generating bases $\bD$. For $\Mt=2$, with
Sylvester-Hadamard construction, we only need to store the Hadamard
matrix, $4$ bits, and $\bD_1$ which has two entries from $2$-bit
quaternary alphabet, for a total of $8$ bits. With the power
construction for $\Mt=2$, each entry of $\bD$ can be stored with $2$-bits,
indicating the sign of the real and imaginary parts. The total storage
required is again $8$ bits. For $\Mt=4$ construction, with the Hadamard
decomposition as in \eqref{eq:MUBdecomposition}, we need $4 \times 2$
bits for the
diagonal matrix (where each quaternary alphabet is represented by $2$
bits) and $4$ bits for the Hadamard matrix (since the
entries are reals only). Therefore, the total required storage is $12$
bits. Note that the MUB codebook storage is independent of $N_b$ and
the same codebook can be used for beamforming and spatial
multiplexing. 

For a fair comparison, \tabref{table:num_bits} shows the number of
bits required to store the Kerdock, Fourier, and Grassmannian codebook
for $\Mt=4$ using $N=16$ for beamforming and $N=8$ for 2-stream
unitary precoded spatial multiplexing. The Kerdock codebook results in 
significant storage savings. 

\subsection{Search Complexity}
For search complexity, we will consider the number of arithmetic
computation required to arrive at the desired codeword. We assume that
\eqref{eq:BF_SC} is tested for beamforming and
\eqref{eq:SM_SC} is tested for spatial multiplexing with the estimated
channel matrix. 
Since the norm computations are common for all codebook entries, we
compare the computation required to compute $\bH \bff$ for
\eqref{eq:BF_SC} and $\bH \bF$ for \eqref{eq:SM_SC} for each codeword
in the codebook. Our search strategy is exhaustive in which the
effective channel gain is computed for all the codewords in the
codebook and the codeword with the largest effective channel gain is
selected as the suboptimal choice. Search space reduction using some
of the well known methods in VQ literature may be possible
\cite{Gersho1991} but we will differ this for future work. In this
paper, we will be concerned with reduction in the arithmetic
computation due to finite alphabet construction of MUB codebooks. 

For the sake of illustration, consider the computation of $\bH \bff$ for the
$2 \times 2$ beamforming case. We want to compute
\begin{eqnarray}
  \bH \bff & = & 
  \begin{bmatrix} h_{11} & h_{12} \\
  h_{21} & h_{22} \end{bmatrix}
  \begin{bmatrix} f_{1} \\ f_{2} \end{bmatrix} \nonumber \\
  & = & 
  \begin{bmatrix} h_{11} f_1 + h_{12} f_2 \\
  h_{21} f_1 + h_{22} f_2 \end{bmatrix}.
\end{eqnarray}
For previously known codebooks, the entries of $\bff$ are complex
valued thus requiring $4$ complex multiplies and $2$ complex
additions. The Kerdock codebook entries are $\{\pm 1,
\pm j\}$ which reduces the complex multiplication into either a sign
change when $f_i = \pm 1$, or swapping the real and imaginary part
with sign change when $f_i = \pm j$. Sign change is a trivial
operation in digital systems and swapping the real and imaginary part
can be accomplished by reading opposite entries in the
memory. Therefore, the Kerdock code effectively achieves
multiplier-less computation of $\bH \bff$ and $\bH \bF$.  

\tabref{table:computation} shows the required number of arithmetic
computation at the receiver for codeword selection. 
For beamforming, the Grassmannian and Fourier based codebooks require
$N \Mt \Mr$ complex multiplies and $N \Mr(\Mt-1)$ complex additions to
find all the candidate effective SNRs. Meanwhile, the proposed Kerdock
codebook does not require any complex multiplication as described
above. Similar elimination of complex multiplication is obtained for
spatial multiplexing. Depending on
the specific implementation for baseband processing, this could
translate to reduced cycle times for complex arithmetic logic
units (ALU) or a possibility for a hard wired logic
implementation. Thus, for a mobile terminals with limited
computational resource and memory, the Kerdock codebook is an
attractive solution for implementation. 

\section{Relationship with Previous Designs}\label{sec:analysis}
In this section, we provide the distance, diversity, and capacity
analysis of the MUB codebook with comparisons to Grassmannian and Fourier
based codebooks. 

\subsection{Distance Properties}
The distance properties are one of the characteristics of MUB
design. The following lemma captures the fact that pairwise chordal distance
between any codewords in the MUB codebook can only take two values. 
\begin{lemma}
For any pair of beamforming MUB codebook elements $\bff_k$ and
$\bff_l$, $k,l=1,2,\dots,N$, in \eqref{eq:KBFCodebook}, the chordal
distance \eqref{eq:chordal_dist} is 
\begin{equation}
  d_{\text{ch}}(\bff_k,\bff_l) = \left\{ \begin{array}{l}
  1 \\ \sqrt{1-\frac{1}{\Mt}}. \end{array} \right.
\end{equation}
\end{lemma}
\begin{IEEEproof}
If $\bff_k$ and $\bff_l$ are from the same ONB, $|\bff_k^* \bff_l|^2 =
0$ since the columns are orthonormal by construction. If $\bff_k$ and
$\bff_l$ are from a different ONB, $|\bff_k^* \bff_l|^2 = 1/\Mt$ by
the mutually unbiased property.
\end{IEEEproof}
It is interesting to observe that as $\Mt \to \infty$, the codebook
approaches that of the Grassmannian codebook. 


For spatial multiplexing, we show that the beamforming Kerdock codebook can be
arranged so that the spatial multiplexing codebook with a large
pairwise chordal distance can be derived. 

Let us consider the proposed spatial multiplexing codebook derived
from the beamforming codebook and examine the projection 2-norm and
Fubini-study distances. First, consider the $\Ms = 2$ spatial
multiplexing codebook for $\Mt = 4$. 
\begin{property}
  Let $\bF_k$ and $\bF_l$, $k \neq l$, be $4 \times 2$ matrices
  composed by taking two columns from any power of $\bD$. Then, 
  \begin{equation}
    |\det (\bF_k^* \bF_l) | = \left\{ \begin{array}{l} 
    0 \\ 1/\sqrt{\Mt}. \end{array} \right.
  \end{equation}
\end{property}
\begin{IEEEproof}
Each $\bF_k$ and $\bF_l$ can be written as $\bF_k = \bD^p \bE_k$ and
$\bF_l = \bD^q \bE_l$ where $\bE_k$ and $\bE_l$ are $4 \times 2$
column selection matrices. Then
\begin{eqnarray}
   \bF_k^* \bF_l & = & \bE_k^T \bD^{q*} \bD^p \bE_l \nonumber \\
   & = & \bE_k^T \bD^r \bE_l, 
\end{eqnarray} 
where $r = (q-p)*$ when $q > p$ and $r = (p-q)$ when $p > q$.
Due to the construction $\bD^r$ is one of the
member bases. The act of $\bE_k^T$ and $\bE_l$ takes $2 \times 2$
submatrix of $\bD^r$. Any
member $\bD^r$ has a structure such that any $2 \times 2$ submatrix
selected this way always contains 1) all reals, 2) a pair of reals and
a pair of imaginary, or 3) all imaginary, from the quaternary
alphabet. It is easy to verify, by listing all possibilities, that the
determinant of such $2 \times 2$ matrix can only
take values $0$ or $1/\sqrt{\Mt}$. 
\end{IEEEproof}

Now consider selecting three columns from each $\bS_n$ to construct a
$\Ms=3$ spatial multiplexing codebook. 
\begin{property}
  Let $\bF_k$ and $\bF_l$, $k \neq l$, be $4 \times 3$ matrices by
  selecting any $3$ columns from each $\bS_n$. Then, 
  \begin{equation}
    |\det (\bF_k^* \bF_l) | = 1/\sqrt{\Mt}.
  \end{equation}
\end{property}
\begin{IEEEproof}
The $\det (\bF_k^*
\bF_l)$ is given by the determinant of a $3 \times 3$ submatrix of
some bases $\bS_n$. Recall that the adjoint of a square matrix $\bD$, denoted
$\text{adj}(\bD)$, is given by $\text{adj}(\bD) = \bD^{-1} \cdot
\det(\bD)$. Since $\bD$ is unitary, $\bD^{-1} = \bD^*$ and $\det(\bD)
= \bI$. So, $\text{adj}(\bD) = \bD^*$. Therefore, the adjoint matrix
also has quaternary alphabet. The elements of adjoint matrix is the
the cofactors which are minors, or determinant of $3 \times 3$
submatrix, with appropriate signs. This shows that every determinant
of $3 \times 3$ submatrix is in the set $\{ \pm 1, \pm j\}$ with
scaling $1/\sqrt{\Mt}$ and the result follows. 
\end{IEEEproof}

Since the projection 2-norm increases as the minimum singular value of 
$\bF_1^* \bF_2$ is decreased, we see that the projection 2-norm is
maximized when we have the column selection such that $|\det (\bF_1^*
\bF_2) | = 0$, or the product matrix is singular. The Fubini-Study
distance \eqref{eq:fs_dist} also maximized by minimizing $|\det
(\bF_1^* \bF_2) |$. 
A $3$-bit Kerdock codebook for $\Ms = 2$ spatial multiplexing in Table \ref{table:sm2}
was found by inspecting the column selections which resulted in the
largest projection 2-norm distance. Note that such subset selection is
not possible with other codebook constructions because the subset
matrix cannot be guaranteed to be a unitary matrix.

\subsection{Diversity}
The diversity order is an important performance metric which indicates
the probability of symbol error trends for high SNR regions. The Kerdock
codebook arranged as in \eqref{eq:KBFCodebook} is easily verified to
have full rank.
\begin{theorem} If $N \geq
  \Mt$, the Kerdock codebook has full diversity order. 
\end{theorem}
\begin{IEEEproof}
The proof follows that found in \cite{Love2003} using the fact that
the Kerdock codebook is of full rank since it is composed of unitary
matrices. Thus, maximum diversity is achieved by the Kerdock
codebook. 
\end{IEEEproof}

\subsection{Capacity}
\begin{figure}[ht]
\centering
\includegraphics[width=5in]{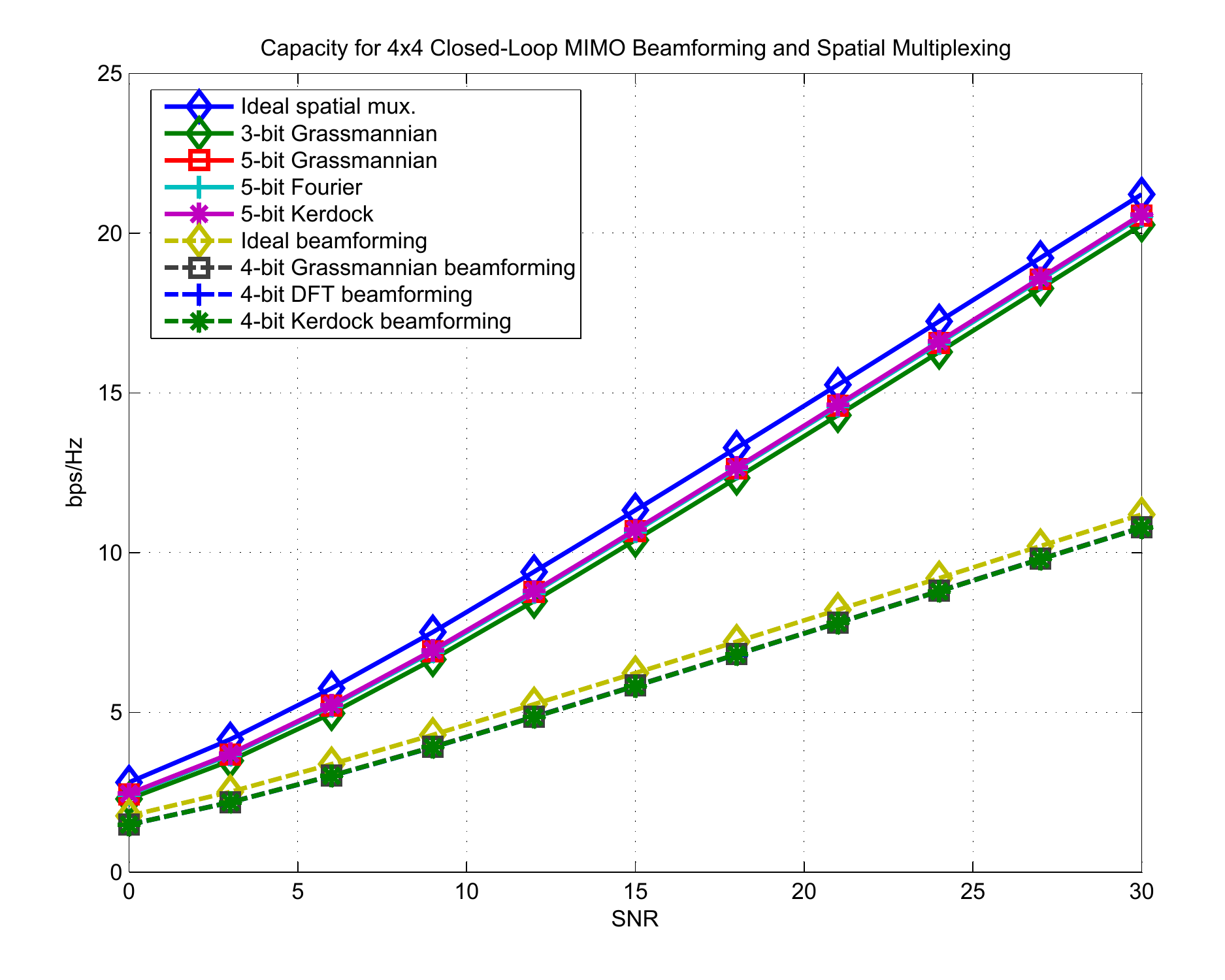} 
\caption{Capacity of $\Mt=\Mr=4$ beamforming system and unitary
  precoded spatial multiplexing system using perfect CSI and the
  Kerdock codebook}
\label{fig:CAP}
\end{figure}

The system capacity associated with quantized codebook is an important
indicator of the quality of the codebook \cite{Love2003,Roh2006a}. The
capacity of the system with a precoder is usually written as
\begin{equation}
  C(\bF) = E_{\bH} \left[ \log_2 \det \left( 
  \bI_{\Ms} + \frac{\cE_s}{\Ms N_o} \bF^* \bH^* \bH \bF
  \right)\right]. 
\end{equation}
where $E_{\bH}$ denotes the expectation with respect to $\bH$. This is
the achievable upper bound when there are no channel estimation errors
and feedback delay, but not the true capacity since power
allocation ({\it i.e.} water filling solution) is not considered. To
give a fair comparison, achievable capacity of
an equal size Grassmannian, Fourier, and Kerdock codebook are
compared in \figref{fig:CAP}. For the
beamforming case (dashed line), we can see that the Grassmannian,
Fourier based, and Kerdock codes have the same achievable capacity
with approximately 1.5dB of loss compared to perfect CSIT. The
spatial multiplexing case (solid line) shows all 5-bit codebooks provide nearly
identical capacity. The 3-bit Grassmannian codebook capacity for
spatial multiplexing system is also shown to illustrate the loss due
to codebook size reduction to 3-bits. Therefore, there is essentially
no capacity loss in using the Kerdock codebook. 

\section{Simulation Results}\label{sec:simulation}
In this section, we give numerical simulation results comparing 
1) Vector Symbol Error Rate (VSER) performance of limited feedback
beamforming system, and 2) VSER performance of two stream unitary
precoded spatial multiplexing system for the Grassmannian, Fourier,
and Kerdock codebook. All simulations are performed for
$\Mt = \Mr = 4$ assuming delay-free feedback. No forward error correction
is used. 

{\it Experiment 1}

\begin{figure}[ht]
\centering
\includegraphics[width=5in]{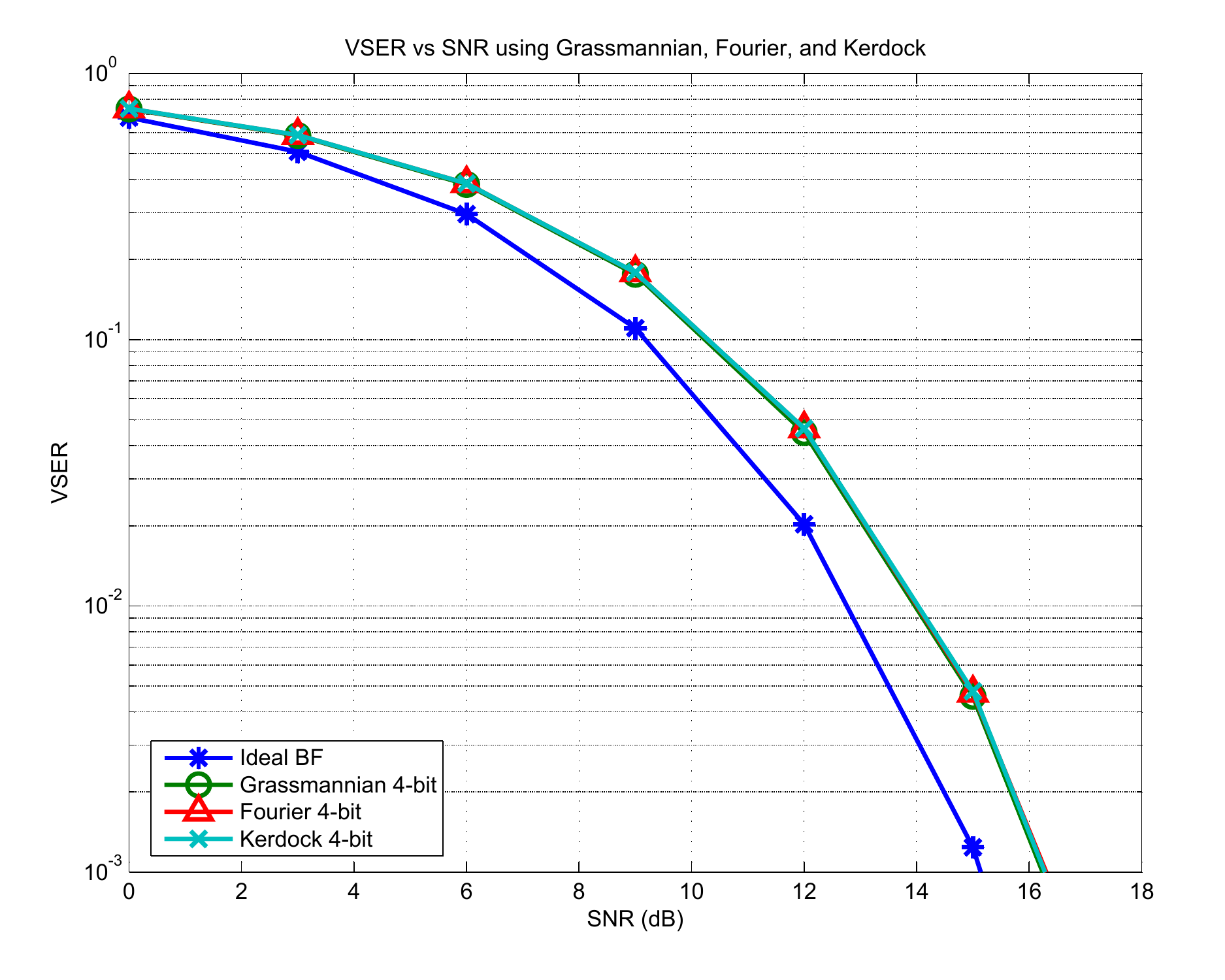}
\caption{Vector Symbol Error Rate performance of $\Mt=\Mr=4$
  beamforming system using 64-QAM comparing perfect CSI, Grassmannian
  codebook, and Kerdock codebook}
\label{fig:VSER_BF}
\end{figure}

The limited feedback beamforming system VSER performance is shown in
\figref{fig:VSER_BF}. For modulation, 64-QAM is used and maximum ratio
combining (MRC) is used at the
receiver. The ideal beamforming result is the lower bound
when perfect CSIT is available. 
The Kerdock codebook outperforms both Grassmannian and Fourier based
codebooks. 

{\it Experiment 2}

\begin{figure}[ht]
\centering
\includegraphics[width=5in]{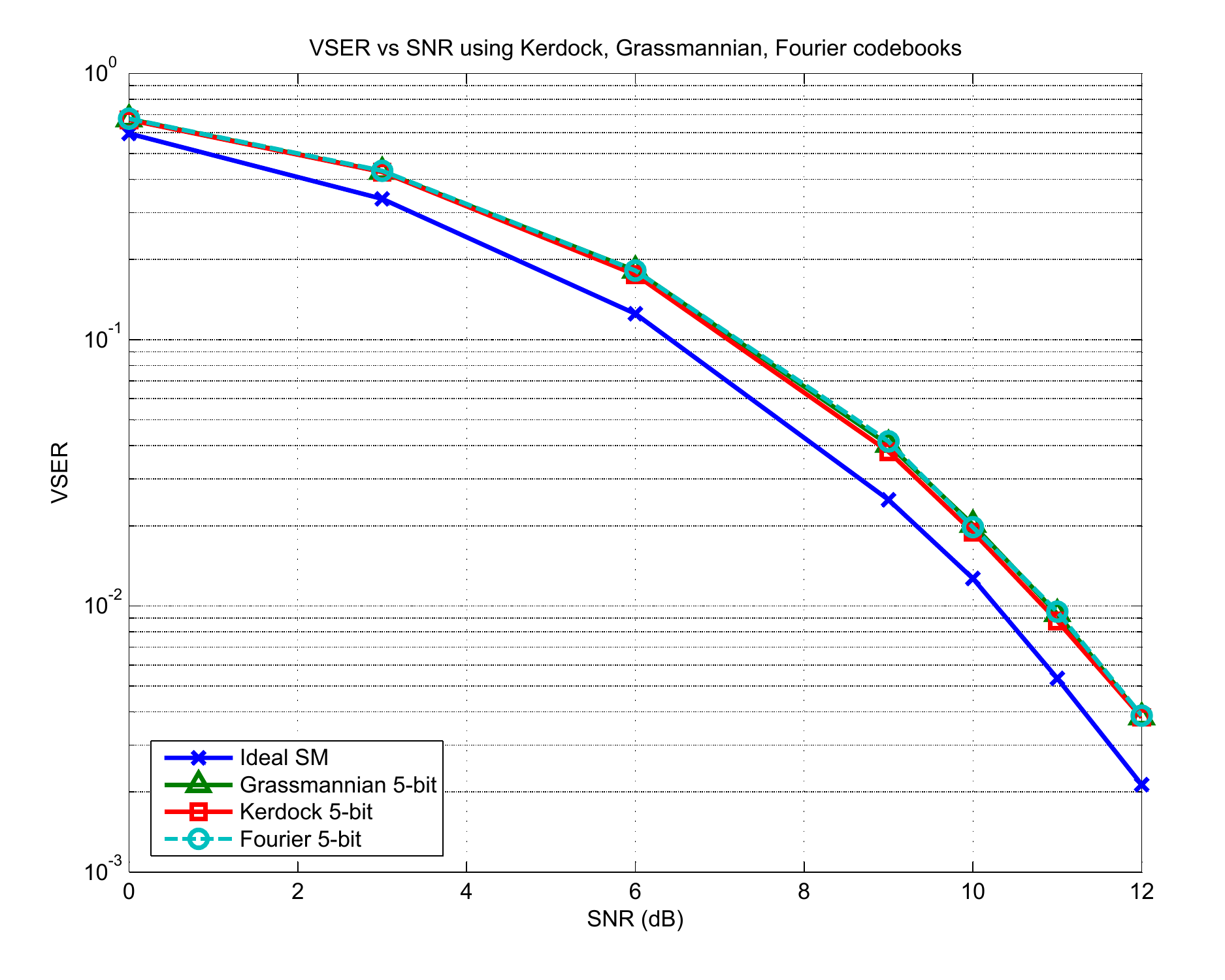}
\caption{Vector Symbol Error Rate Performance of $\Mt=\Mr=4$ Spatial
  Multiplexing System 16-QAM comparing perfect CSI, Grassmannian
  codebook, and Kerdock codebook}
\label{fig:VSER_SM}
\end{figure}

The VSER performance for limited feedback two-stream unitary precoded
spatial multiplexing system using 5-bit codebook is shown in
\figref{fig:VSER_SM}. A 16-QAM modulation and zero-forcing receiver is
used. The
Kerdock codebook outperforms the Grassmannian
codebook and the Fourier based codebook, despite having only $30$
entries. 

\begin{figure}[ht]
\centering
\includegraphics[width=5in]{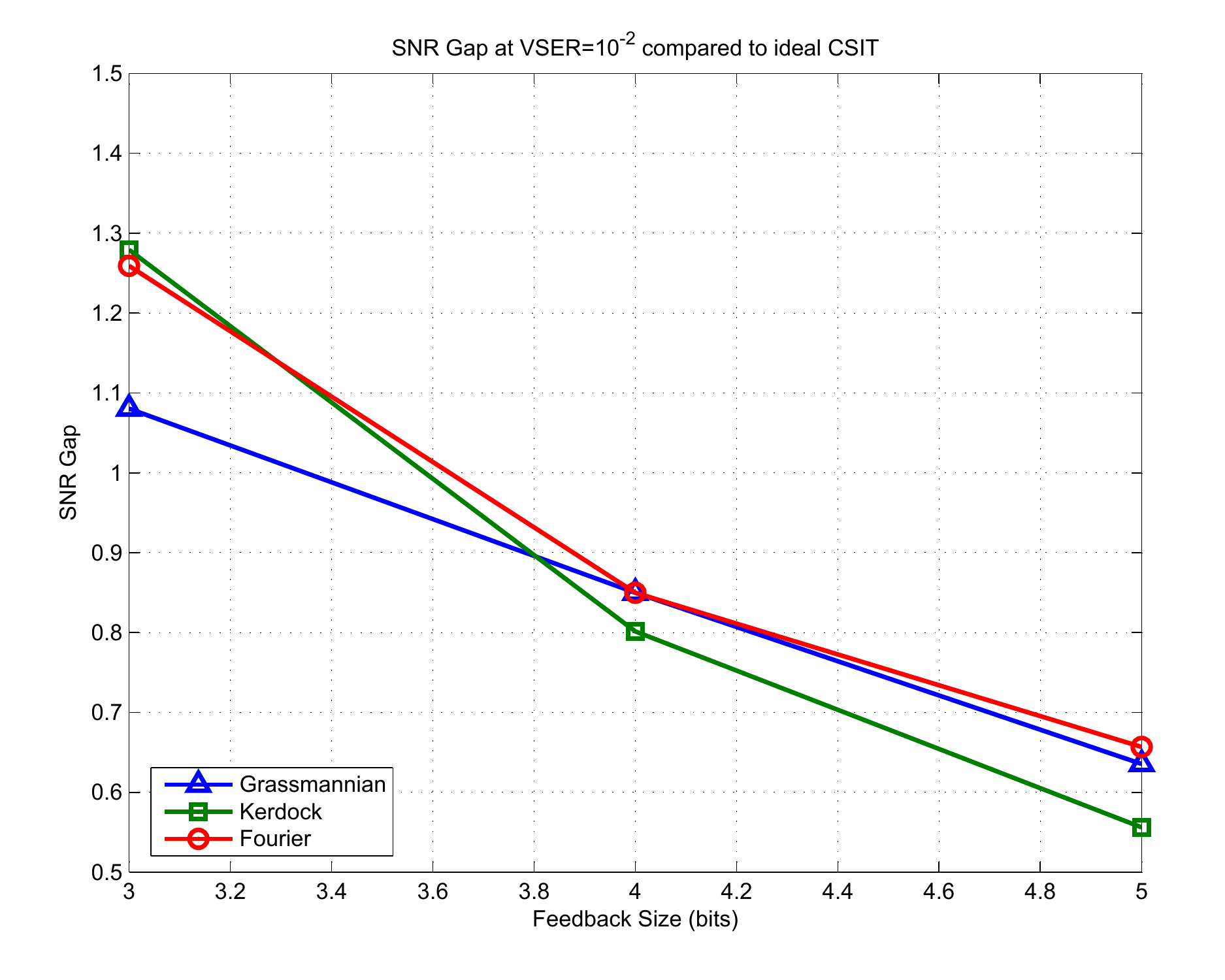}
\caption{SNR gap between ideal CSI case and various codebook design
  for 2-stream Spatial Multiplexing System at VSER = $10^{-2}$}
\label{fig:SM_SNRgap}
\end{figure}

To clearly see the performance difference among the codebook designs,
\figref{fig:SM_SNRgap} shows the SNR gap between the ideal CSIT case and
the three codebook designs at ${\text VSER} = 10^{-2}$. As expected,
the Grassmannian codebook outperforms the Fourier codebook. The
Kerdock codebook shows worse performance for $3$-bit codebook because
only $8$ of $30$ possible codewords are used. However, as we increase
the codebook size to $4$ and $5$ bits, the Kerdock codebook
outperforms the Grassmannian codebook which is quite remarkable
considering the fact that the codebook contains only quaternary alphabet. 

Overall, the results indicate that the Kerdock codebook can perform
comparable or better than previously known codebooks with additional
benefit of 1) structured construction, 2) finite alphabet, 3) reduced
search complexity, and 4) shared codebook between beamforming and spatial multiplexing. 

\section{Conclusion}\label{sec:conclusion}
In this paper, we proposed to use Kerdock codes for limited feedback
precoded MIMO systems. The Kerdock code is an MUB set
with correlation properties that can be linked to Grassmannian line
packing problem, equiangular frames, and Welch bound equality sequence
sets. We showed that the Kerdock codes can achieve full diversity and
performance comparable or better than previously known codebooks with
additional benefits of finite alphabet construction, reduced storage
and search requirements, and shared codebook between beamforming and
spatial multiplexing. We found that there is essentially no loss in
achievable capacity compared to equal size Grassmannian and Fourier
based codebooks. One limitation of this work is that the Kerdock
codes can only be constructed for number of transmit antennas which
are power of two. An open problem remains in constructing odd
dimension codebook with finite alphabet entries. Our future work will
consider effects of space or time correlated channels and possible extensions
to multiuser scenarios. In particular, Kerdock codes are possibly
applicable to a multiuser MIMO system using a unitary basis sets,
known as PU$^2$RC \cite{Samsung2006,Huang2007}.  

\ifCLASSOPTIONcaptionsoff
  \newpage
\fi



%

\bibliographystyle{IEEEtran}

\begin{thebibliography}{10}
\providecommand{\url}[1]{#1}
\csname url@samestyle\endcsname
\providecommand{\newblock}{\relax}
\providecommand{\bibinfo}[2]{#2}
\providecommand{\BIBentrySTDinterwordspacing}{\spaceskip=0pt\relax}
\providecommand{\BIBentryALTinterwordstretchfactor}{4}
\providecommand{\BIBentryALTinterwordspacing}{\spaceskip=\fontdimen2\font plus
\BIBentryALTinterwordstretchfactor\fontdimen3\font minus
  \fontdimen4\font\relax}
\providecommand{\BIBforeignlanguage}[2]{{%
\expandafter\ifx\csname l@#1\endcsname\relax
\typeout{** WARNING: IEEEtran.bst: No hyphenation pattern has been}%
\typeout{** loaded for the language `#1'. Using the pattern for}%
\typeout{** the default language instead.}%
\else
\language=\csname l@#1\endcsname
\fi
#2}}
\providecommand{\BIBdecl}{\relax}
\BIBdecl

\bibitem{Goldsmith2003}
A.~Goldsmith, S.~Jafar, N.~Jindal, and S.~Vishwanath, ``Capacity limits of
  {MIMO} channels,'' \emph{{IEEE} J. Sel. Areas Commun.}, vol.~21, no.~5, pp.
  684--702, 2003.

\bibitem{Narula1998}
A.~Narula, M.~Lopez, M.~Trott, and G.~Wornell, ``Efficient use of side
  information in multiple-antenna data transmission over fading channels,''
  \emph{{IEEE} J. Sel. Areas Commun.}, vol.~16, no.~8, pp. 1423--1436, 1998.

\bibitem{Mukkavilli2003}
K.~Mukkavilli, A.~Sabharwal, E.~Erkip, and B.~Aazhang, ``On beamforming with
  finite rate feedback in multiple-antenna systems,'' \emph{{IEEE} Trans. Inf.
  Theory}, vol.~49, no.~10, pp. 2562--2579, 2003.

\bibitem{Love2003}
D.~J. Love, R.~W. Heath~Jr, and T.~Strohmer, ``Grassmannian beamforming for
  multiple-input multiple-output wireless systems,'' \emph{{IEEE} Trans. Inf.
  Theory}, vol.~49, no.~10, pp. 2735--2747, 2003.

\bibitem{Roh2004}
J.~Roh and B.~Rao, ``Performance analysis of multiple antenna systems with
  {VQ}-based feedback,'' in \emph{Signals, Systems and Computers, 2004.
  Conference Record of the Thirty-Eighth Asilomar Conference on}, vol.~2, 2004,
  pp. 1978--1982 Vol.2.

\bibitem{Love2005a}
D.~J. Love and R.~W. Heath~Jr, ``Limited feedback unitary precoding for spatial
  multiplexing systems,'' \emph{{IEEE} Trans. Inf. Theory}, vol.~51, no.~8, pp.
  2967--2976, 2005.

\bibitem{Mondal2006a}
B.~Mondal and R.~W. Heath~Jr, ``Performance analysis of quantized beamforming
  {MIMO} systems,'' \emph{{IEEE} Trans. Signal Process.}, vol.~54, no.~12, pp.
  4753--4766, 2006.

\bibitem{3GPP}
\BIBentryALTinterwordspacing
{3GPP}, ``Physical layer aspects of {UTRA} high speed downlink packet access,''
  \emph{Technical Report TR25.814}, 2006. [Online]. Available:
  \url{http://www.3gpp.org/ftp/Specs/html-info/Meetings-R1.htm}
\BIBentrySTDinterwordspacing

\bibitem{Ryan2007a}
D.~J. Ryan, I.~V.~L. Clarkson, I.~B. Collings, D.~Guo, and M.~L. Honig, ``{QAM}
  codebooks for low-complexity limited feedback {MIMO} beamforming,'' in
  \emph{Proc. IEEE Intl. Conf. on Comm.}, 2007, pp. 4162--4167.

\bibitem{Gersho1991}
A.~Gersho and R.~M. Gray, \emph{Vector Quantization and Signal
  Compression}.\hskip 1em plus 0.5em minus 0.4em\relax Kluwer Academic, 1991.

\bibitem{Roh2006}
J.~Roh and B.~Rao, ``Transmit beamforming in multiple-antenna systems with
  finite rate feedback: a {VQ}-based approach,'' \emph{{IEEE} Trans. Inf.
  Theory}, vol.~52, no.~3, pp. 1101--1112, 2006.

\bibitem{Roh2006a}
------, ``Design and analysis of {MIMO} spatial multiplexing systems with
  quantized feedback,'' \emph{{IEEE} Trans. Signal Process.}, vol.~54, no.~8,
  pp. 2874--2886, 2006.

\bibitem{Xia2006}
P.~Xia and G.~Giannakis, ``Design and analysis of transmit-beamforming based on
  limited-rate feedback,'' \emph{{IEEE} Trans. Signal Process.}, vol.~54,
  no.~5, pp. 1853--1863, 2006.

\bibitem{Zheng2007}
J.~Zheng, E.~R. Duni, and B.~D. Rao, ``Analysis of multiple-antenna systems
  with finite-rate feedback using high-resolution quantization theory,''
  \emph{{IEEE} Trans. Signal Process.}, vol.~55, no.~4, pp. 1461--1476, 2007.

\bibitem{Lau2004}
V.~Lau, Y.~Liu, and T.-A. Chen, ``On the design of {MIMO} block-fading channels
  with feedback-link capacity constraint,'' \emph{{IEEE} Trans. Commun.},
  vol.~52, no.~1, pp. 62--70, 2004.

\bibitem{Santipach2004}
W.~Santipach and M.~Honig, ``Asymptotic capacity of beamforming with limited
  feedback,'' in \emph{Information Theory, 2004. ISIT 2004. Proceedings.
  International Symposium on}, 2004, pp. 290--.

\bibitem{Zhou2005}
S.~Zhou, Z.~Wang, and G.~Giannakis, ``Quantifying the power loss when transmit
  beamforming relies on finite-rate feedback,'' \emph{{IEEE} Trans. Wireless
  Commun.}, vol.~4, no.~4, pp. 1948--1957, 2005.

\bibitem{Mondal2007}
B.~Mondal, S.~Dutta, and R.~W. Heath~Jr, ``Quantization on the grassmann
  manifold,'' \emph{{IEEE} Trans. Signal Process.}, vol.~55, no.~8, pp.
  4208--4216, 2007.

\bibitem{Conway1996}
J.~H. Conway, R.~H. Hardin, and N.~J.~A. Sloane, ``Packing lines, planes, etc.:
  packings in grassmannian spaces,'' \emph{Experimental Mathematics}, vol.~5,
  pp. 139--159, 1996.

\bibitem{Shor1998}
P.~Shor and N.~Sloane, ``A family of optimal packings in grassmannian
  manifolds,'' \emph{Journal of Algebraic Combinatorics}, vol.~7, no.~2, pp.
  157--163, Mar. 1998.

\bibitem{Bachoc2006}
C.~Bachoc, ``Linear programming bounds for codes in grassmannian spaces,''
  \emph{{IEEE} Trans. Inf. Theory}, vol.~52, no.~5, pp. 2111--2125, 2006.

\bibitem{Barg2006}
A.~Barg and D.~Nogin, ``A bound on grassmannian codes,'' in \emph{Information
  Theory, 2006 IEEE International Symposium on}, 2006, pp. 997--1000.

\bibitem{Tropp2006}
J.~A. Tropp, I.~S. Dhillon, R.~W. Heath~Jr, and T.~Strohmer, ``Constructing
  packings in grassmannian manifolds via alternating projections,''
  \emph{Submitted to Experimental Mathematics}, Nov. 2006.

\bibitem{Mondal2005a}
B.~Mondal, R.~Samanta, and R.~W. Heath~Jr, ``Frame theoretic quantization for
  limited feedback {MIMO} beamforming systems,'' in \emph{Proc. 2005 Int. Conf.
  on Wireless Networks, Communications and Mobile Computing}, vol.~2, 2005, pp.
  1065--1070 vol.2.

\bibitem{Strohmer2003}
T.~Strohmer and R.~W. Heath~Jr, ``Grassmannian frames with applications to
  coding and communication,'' \emph{Applied and Computational Harmonic
  Analysis}, vol.~14, no.~3, pp. 257--275, May 2003.

\bibitem{Tropp2005}
J.~A. Tropp, I.~S. Dhillon, R.~W. Heath~Jr, and T.~Strohmer, ``Designing
  structured tight frames via an alternating projection method,'' \emph{{IEEE}
  Trans. Inf. Theory}, vol.~51, no.~1, pp. 188--209, 2005.

\bibitem{Hochwald2000}
B.~Hochwald, T.~Marzetta, T.~Richardson, W.~Sweldens, and R.~Urbanke,
  ``Systematic design of unitary space-time constellations,'' \emph{{IEEE}
  Trans. Inf. Theory}, vol.~46, no.~6, pp. 1962--1973, 2000.

\bibitem{Welch1974}
L.~Welch, ``Lower bounds on the maximum cross correlation of signals
  (corresp.),'' \emph{{IEEE} Trans. Inf. Theory}, vol.~20, no.~3, pp. 397--399,
  1974.

\bibitem{Xia2005}
P.~Xia, S.~Zhou, and G.~Giannakis, ``Achieving the welch bound with difference
  sets,'' \emph{{IEEE} Trans. Inf. Theory}, vol.~51, no.~5, pp. 1900--1907,
  2005.

\bibitem{Ryan2006}
D.~J. Ryan, I.~B. Collings, I.~Vaughan, and L.~Clarkson, ``Maximum-likelihood
  noncoherent lattice decoding of {QAM},'' in \emph{Proc. IEEE Int. Conf. on
  Acoustics, Speech and Signal Processing}, vol.~4, 2006, pp. IV--IV.

\bibitem{Kerdock1972}
A.~Kerdock, ``Studies of low-rate binary codes (ph.d. thesis abstr.),''
  \emph{{IEEE} Trans. Inf. Theory}, vol.~18, no.~2, pp. 316--316, 1972.

\bibitem{Hammons1994}
A.~Hammons~Jr., P.~Kumar, A.~Calderbank, N.~Sloane, and P.~Sole, ``The
  {Z4}-linearity of kerdock, preparata, goethals, and related codes,''
  \emph{{IEEE} Trans. Inf. Theory}, vol.~40, no.~2, pp. 301--319, 1994.

\bibitem{Heath2006}
R.~W. Heath~Jr, T.~Strohmer, and A.~J. Paulraj, ``On quasi-orthogonal
  signatures for cdma systems,'' \emph{{IEEE} Trans. Inf. Theory}, vol.~52,
  no.~3, pp. 1217--1226, 2006.

\bibitem{Heath2001}
R.~W. Heath~Jr, S.~Sandhu, and A.~J. Paulraj, ``Antenna selection for spatial
  multiplexing systems with linear receivers,'' \emph{{IEEE} Commun. Lett.},
  vol.~5, no.~4, pp. 142--144, 2001.

\bibitem{Gow2007}
\BIBentryALTinterwordspacing
R.~Gow, ``Generation of mutually unbiased bases as powers of a unitary matrix
  in 2-power dimensions,'' 2007. [Online]. Available:
  \url{http://arxiv.org/abs/math/0703333}
\BIBentrySTDinterwordspacing

\bibitem{Spencer2004a}
Q.~Spencer, C.~Peel, A.~Swindlehurst, and M.~Haardt, ``An introduction to the
  multi-user {MIMO} downlink,'' \emph{{IEEE} Commun. Mag.}, vol.~42, no.~10,
  pp. 60--67, 2004.

\bibitem{Choi2005}
J.~Choi and R.~W. Heath~Jr, ``Interpolation based transmit beamforming for
  {MIMO-OFDM} with limited feedback,'' \emph{{IEEE} Trans. Signal Process.},
  vol.~53, no.~11, pp. 4125--4135, 2005.

\bibitem{Mondal2006c}
B.~Mondal and R.~W. Heath~Jr, ``On the {SNR} and diversity of quantized
  precoded {MIMO} systems,'' in \emph{Proc. 7th IEEE Workshop on Signal
  Processing Advances in Wireless Communications (SPAWC 2006)}, 2006, pp. 1--5.

\bibitem{Klappenecker2005}
A.~Klappenecker and M.~Roetteler, ``Mutually unbiased bases, spherical designs,
  and frames,'' in \emph{Proceedings of SPIE}, M.~Papadakis, A.~F. Laine, and
  M.~A. Unser, Eds., vol. 5914, no.~1.\hskip 1em plus 0.5em minus 0.4em\relax
  SPIE, 2005, p. 59140.

\bibitem{Alltop1980}
W.~Alltop, ``Complex sequences with low periodic correlations (corresp.),''
  \emph{{IEEE} Trans. Inf. Theory}, vol.~26, no.~3, pp. 350--354, 1980.

\bibitem{Wootters1989}
W.~K. Wootters and B.~D. Fields, ``Optimal state-determination by mutually
  unbiased measurements,'' \emph{Annals of Physics}, vol. 191, no.~2, pp.
  363--381, May 1989.

\bibitem{Klappenecker2004}
A.~Klappenecker and M.~Roetteler, ``Constructions of mutually unbiased bases,''
  \emph{Finite Fields and Applications}, pp. 137--144, 2004.

\bibitem{Bandyopadhyay2002}
S.~Bandyopadhyay, P.~O. Boykin, V.~Roychowdhury, and F.~Vatan, ``A new proof
  for the existence of mutually unbiased bases,'' \emph{Algorithmica}, vol.~34,
  no.~4, pp. 512--528, Nov. 2002.

\bibitem{Planat2006}
M.~R.~P. Planat, H.~Rosu, S.~Perrine, and M.~Saniga, ``A survey of finite
  algebraic geometrical structures underlying mutually unbiased quantum
  measurements,'' \emph{Foundations of Physics}, vol.~36, p. 1662, 2006.

\bibitem{Nilsson2005}
A.~Nilsson, E.~Tell, D.~Wiklund, and D.~Liu, ``Design methodology for
  memory-efficient multi-standard baseband processors,'' in \emph{Proc. 2005
  Asia-Pacific Conference on Communications}, 2005, pp. 28--32.

\bibitem{Markovic2007}
D.~Markovic, B.~Nikolic, and R.~Brodersen, ``Power and area minimization for
  multidimensional signal processing,'' \emph{{IEEE} J. Solid-State Circuits},
  vol.~42, no.~4, pp. 922--934, 2007.

\bibitem{LoveCodebook}
\BIBentryALTinterwordspacing
D.~J. Love, ``Grassmannian subspace packing webpage.'' [Online]. Available:
  \url{http://cobweb.ecn.purdue.edu/~djlove/grass.html}
\BIBentrySTDinterwordspacing

\bibitem{Samsung2006}
\BIBentryALTinterwordspacing
Samsung, ``Downlink {MIMO} for {EUTRA},'' \emph{{3GPP TSG WG1} Meeting 44,
  R1-060335}, 2006. [Online]. Available:
  \url{http://www.3gpp.org/ftp/tsg_ran/WG1_RL1/TSGR1_44/Docs/R1-060335.zip}
\BIBentrySTDinterwordspacing

\bibitem{Huang2007}
\BIBentryALTinterwordspacing
K.~Huang, J.~G. Andrews, and R.~W. Heath~Jr, ``Performance of orthogonal
  beamforming for sdma with limited feedback,'' \emph{submitted to IEEE Trans.
  on Vehicular Technology}, 2007. [Online]. Available:
  \url{http://users.ece.utexas.edu/~khuang/Papers/TVT07/}
\BIBentrySTDinterwordspacing

\end{thebibliography}

%






\begin{table}[htbp]
  \centering
  \renewcommand{\arraystretch}{1.3}
  \caption{Unitary precoding codebook for 2 streams derived from the
  beamforming codebook. The column selection in the subscript is
  chosen to maximize the pairwise projection 2-norm distance.}
  \label{table:sm2}
  \begin{tabular}{|c|c|c|c|}
    \hline
    $[\bS_0]_{1,2}$ & $[\bS_0]_{3,4}$ & $[\bS_1]_{1,3}$ &
    $[\bS_1]_{2,4}$ \\ \hline
    $[\bS_2]_{1,4}$ & $[\bS_2]_{2,3}$ & $[\bS_3]_{1,4}$ &
    $[\bS_3]_{2,3}$ \\ \hline
  \end{tabular}
\end{table}

\begin{table}
  \centering
  \renewcommand{\arraystretch}{1.3}
  \caption{Number of bits required for storing MUB, Fourier,
  and Grassmannian codebooks for $\Mt=4$ and using $N=16$
  for beamforming and $N=8$ for 2-stream spatial multiplexing.
  A system dependent number of bits used to represent a real number is denoted by $N_b$.}
  \label{table:num_bits}
  \begin{tabular}{|c|c|c|}
    \hline
    MUB & Fourier & Grassmannian \\ \hline
    $12$ & $40N_b$ & $256N_b$ \\ \hline
  \end{tabular}
\end{table}

\begin{table}
  \centering
  \renewcommand{\arraystretch}{1.3}
  \caption{Comparison of computational requirement for codeword
  selection}
  \label{table:computation}
  \begin{tabular}{|c|c|c|}
    \hline
    \multicolumn{3}{|c|}{Beamforming Selection} \\ \hline
    & Grassmannian or Fourier & Kerdock \\ \hline
    Multiply & $N \Mt \Mr$ & 0  \\ \hline
    Addition  & $N \Mr (\Mt-1)$ & $N \Mr (\Mt-1)$ \\ \hline \hline
    \multicolumn{3}{|c|}{Spatial Multiplexing: Projection 2-norm} \\ \hline
    & Grassmannian or Fourier & Kerdock \\ \hline
    Multiply & $N \Ms \Mr^2$ & 0 \\ \hline
    Addition  & $N \Mr^2(\Ms-1)$ & $N \Mr^2(\Ms-1)$ \\ \hline \hline
    \multicolumn{3}{|c|}{Spatial Multiplexing: Fubini-Study} \\ \hline
    & Grassmannian or Fourier & Kerdock \\ \hline
    Multiply & $N \Ms^2 \Mr$ & 0 \\ \hline
    Addition  & $N \Ms^2(\Mr-1)$ & $N \Ms^2(\Mr-1)$ \\ \hline
  \end{tabular}
\end{table}

\end{document}